**Lipids and lipid-mixtures in boundary layers: from hydration lubrication to osteoarthritis**

Yifeng Cao[a,*], Jacob Klein*

Department of Molecular Chemistry and Materials Science, Weizmann Institute of Science, Rehovot 76100, Israel

* caoyf@zju.edu.cn (Y. C.); jacob.klein@weizmann.ac.il (J. K.)

[a] Current address: Institute of Zhejiang University-Quzhou, Quzhou 324000, China

**Abstract**

The hydration layer surrounding the phosphocholine headgroups of single-component phosphatidylcholine (PC) lipids, or of lipid-mixtures, assembled at an interface greatly modifies the interfacial properties and interactions. As water molecules within the hydration layer are held tightly by the headgroup but are nonetheless very fluid upon shear, the boundary lipid layers, exposing the highly hydrated headgroup arrays, can provide efficient boundary lubrication when sliding against an opposing surface, at physiologically-high contact pressures. Additionally, any free lipids in the surrounding liquid can heal defects which may form during sliding on the boundary PC layer. Similar boundary lipid layers contribute to the lubricating, pressure-bearing, and wear-protection functions of healthy articular joints. This review presents a survey of the relationship between the molecular composition of the interfacial complex and the lubrication behavior of the lipid-based boundary layers, which could be beneficial for designing boundary lubricants for intra-articular injection for the treatment of early OA.

## 1. Introduction

The major articular joints of mammals, such as hips and knees, display extraordinarily efficient lubrication at the articular cartilage coating the joints, as well as high-pressure bearing and self-healing lubricating layers, which enables life-long movements of the joints and the body [1,2]. Synovial joints are composed of opposing articular cartilage layers coating the joint ends, and surrounded by synovial fluid within the joint capsule. When the opposing articular cartilage surfaces slide pass each other, the friction coefficient $\mu$, defined as $\mu =$ [(force required to slide)/load compressing the surfaces)], has very low values, down to $\mu \approx 0.03 - 0.001$ under a wide range of contact pressures up to 5 MPa or even higher [1,3,4]. Such an extent of friction reduction is better than can be achieved by any artificial materials, and is indeed crucial for the homeostasis of cartilage in joints [5]. Such remarkable lubrication behavior has been attributed to a number of factors, including fluid-film lubrication by synovial fluid, interstitial fluid pressurization for load support, and boundary lubrication under high pressures [5–7].

Under the boundary lubrication regime, which is increasingly believed to be the main mechanism responsible for the low cartilage friction [5,8–10], friction arises due to energy dissipation when the directly-contacting opposing boundary layers on the cartilage surfaces slide past each other. Thus, the properties and lubrication behavior of molecules as well as their assemblies constructing this boundary layer plays a dominant role in synovial joint lubrication. Components of the synovial fluid surrounding the cartilage, as well as components within the cartilage layer itself that are produced by the chondrocyte cells embedded in it and migrate to the surface, are the reservoir providing molecules that renew this boundary layer when it undergoes wear (as it must) [11]. On the one hand, molecules in synovial fluid may directly adsorb onto the cartilage surface; on the other hand, synovial fluid provides nutrients to the underlying cartilage during increase and decrease of normal load. Small molecules from synovial fluid penetrate through the interfacial cartilage layer and provide nutrients to chondrocytes embedded in the deeper cartilage zone, while metabolites of chondrocytes are transferred in the other direction [12,13].

Studies have shown that macromolecular components at the cartilage surface, including hyaluronic acid (HA, also known as hyaluronan), proteoglycan (mostly lubricin on the surface), and the collagen network itself, assemble by intermolecular interactions and are essential for maintaining the lubrication ability of the joints [5,14]. Increasing evidence, with the exception of one recent study on lubricin [15], suggest that biomacromolecules, such as HA, aggrecan, and lubricin, cannot provide as efficient lubrication under high load as that in synovial joints [16–18]. Rather, surface-active phospholipids (PLs) lining the cartilage surface have been suggested as the key components responsible for the lubrication (Figure 1) [5,19]. HA complexed with PLs can offer superior boundary lubrication by exposing a hydrated PL headgroup layer at the interface, attributed to the hydration lubrication mechanism [10,19-21], while HA molecules may act as supporting matrix for the boundary PL layer [22]. Together with lubricin molecules, which are proposed to bind HA to the cartilage, the supramolecular, lipid-based superficial layer may account for the boundary lubrication of articular joints [5,8].

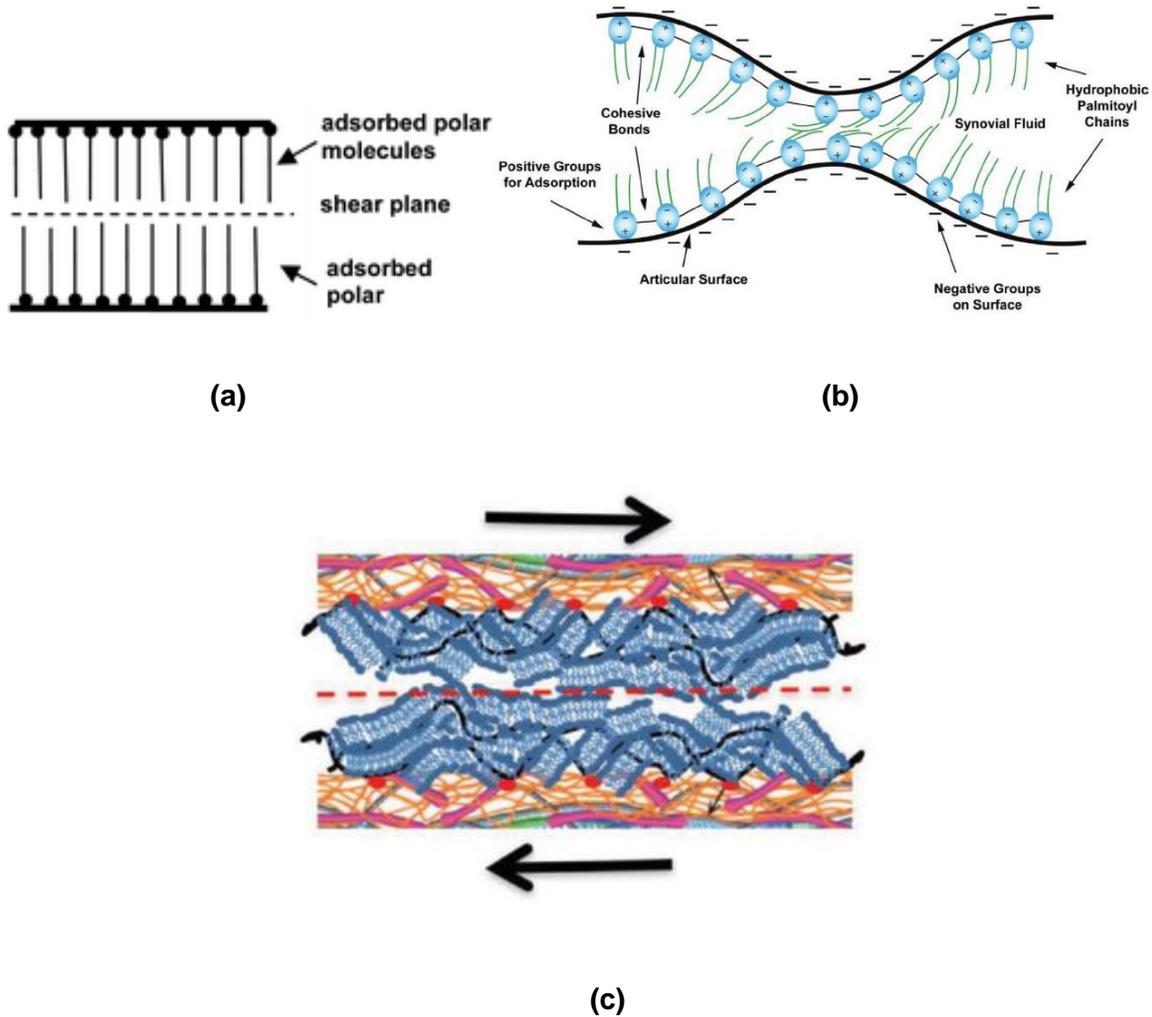

**(a)**

**(b)**

**(c)**

**Figure 1.** Models of boundary lubrication. (a) Early model of boundary lubrication by monolayer of polar molecules adsorbed on the substrate, proposed by Hardy et. al. in 1922 [23]. In this picture, sliding at the slip-plane involves energy dissipation through breakage of van der Waals bonds between alkyl tails of the surfactants, which, while quite high and associated with friction coefficients of order 0.1, protects the underlying substrate from higher friction and wear. (b) A model proposed by Hills [24] for boundary lubrication between cartilage surfaces, conceptually similar to that of Hardy in (a), where boundary lubrication is via slip at the tail-tail interface of opposing surface active phospholipids in synovial joint. Phospholipids adsorb on the negatively charged (cartilage) surface by their zwitterionic headgroups and expose the hydrophobic tails to the interface. Friction coefficients for such a configuration are also expected to be of order 0.1. (c) Boundary lubrication by phospholipids boundary layers complexed with macromolecules, such as HA and lubricin, on the cartilage surface [5][19]. Friction coefficients down to $10^{-3}$ up to physiological pressures are generated at the hydrated

headgroup-headgroup interface. For panel (a), reproduced with permission from Ref. [7]; Copyright © 2013 Royal Society of Chemistry. For panels (b) and (c), reproduced with permission from Ref. [7]; Copyright © 2015, American Chemical Society.

The presence of PLs lining the cartilage surface and present in synovial fluid has been confirmed both visually by electron microscopy and analytically by mass spectrometry combined with liquid or gas chromatography [25–28]. These PLs may be mostly synthesized by fibroblast-like synoviocytes, a specialized cell type located in the synovium inside the joints, and partially from blood [29]. A complicated PL mixture composed of more than a hundred PLs belonging to several groups: phosphatidylcholine (PC) is the most-abundant PL group found both on the cartilage surface (41 %) and in synovial fluid (67 %); sphingomyelin (SM), phosphatidylethanolamine (PE), as well as negatively-charged PLs, such as phosphatidic acid (PA) and phosphatidylglycerol (PG), have also been detected in synovial joints [27,28,30]. The majority of the PLs in synovial joints are unsaturated ones, particularly, only 5.4 % of PCs in synovial fluid are saturated [28]. The chemical structure of PL, not only determines the physiochemical properties of the lipid and lipid assembly, but also dominates its interaction with macromolecular matrix and surrounding ions, which in turn influences the boundary lubrication efficiency [27]. Additionally, the existence of PL mixtures in cartilage and in synovial joints may suggest some evolutionary advantages in the bio-functions of the interfacial layer, including lubrication, pressure-bearing, and self-healing; but this is not clearly understood to date.

The contents and composition of PLs, as well as ratios between certain PLs and PL groups, are also associated with joint diseases, rheumatoid arthritis (RA) and osteoarthritis (OA) [30,31]. OA is the most common degenerative joint disease worldwide. It is characterized by cartilage destruction, subchondral bone sclerosis, and osteophyte formation [32], affecting millions of people, primarily elderly. Early OA initiates the degradation of articular cartilage, increases the friction coefficient, and causes progressive damage. Therefore, improving lubrication with PL boundary layer may offer a feasible option for the treatment of early OA [33].

In the context of boundary lubrication, the sliding surfaces are in molecular contact with each other and frictional energy dissipation arises from interactions between superficial PL layers. Herein, we will address the current and emerging results in utilizing interfacial lipid(s) layers as efficient boundary lubricants by starting with hydration lubrication − the origin of extreme lubrication by hydrated species; following that, we will summarize the boundary lubrication behavior of single-component lipid layer and put effort on factors influencing the lubrication efficiency; after that, we will discuss how multi-component PLs layer behave as boundary lubricants, whereas emphasis will be put on the differences between single- and multiple-PL(s) as well as the implication to synovial joint.

## 2. Hydration lubrication

Direct measurement of the normal pressure-distance relationship using either the osmotic stress method or surface force balance/apparatus (SFB or SFA) [34], for respective free-standing and supported PL bilayers, shows that [35–37]: (1) a hydration layer of a few nanometers thick is trapped between PL bilayers (PC, PE, and SM bilayers); (2) a strong, short-ranged repulsion is the dominant force acting between PL bilayers when they are in close contact (inter-bilayer separation $D$ < hydration radius, ca. 2 nm); (3) at high pressures, the repulsive force decays exponentially with a decay length of a few angstroms. Considering these PL headgroups are zwitterionic and highly-hydrated [38–40], such strong repulsive interaction is likely to be due to hydration repulsion [41]. In aqueous environments water molecules interact with the zwitterionic headgroups via strong dipole-charge interactions (Figure 2), giving rise to a hydration layer surrounding the headgroups; when the bilayers are in contact, strong energy is required to remove water molecules from the hydration layer, i.e. dehydration of the headgroups [42].

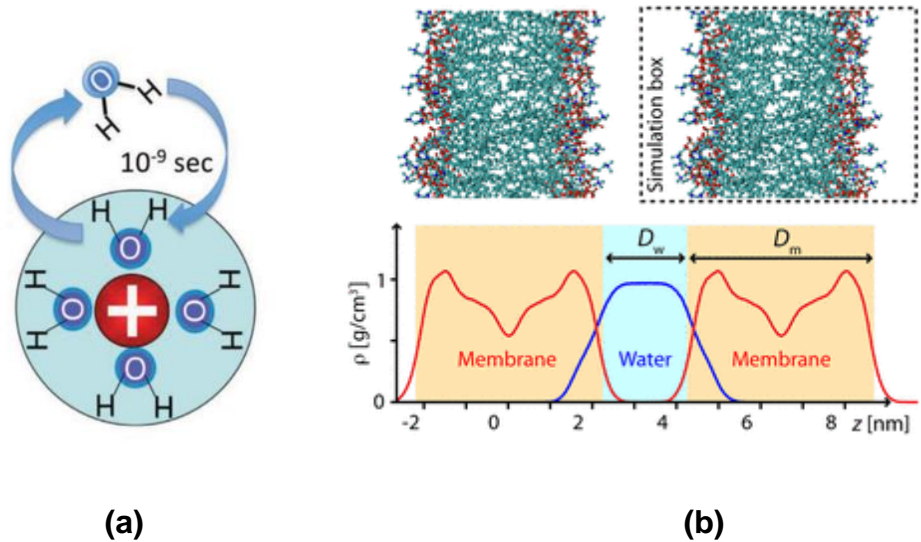

**(a)**                    **(b)**

**Figure 2**. (a) Schematic of the hydration shell surrounding a charge [42]. Water molecules within the shell are strongly attached to the charge via charge-dipole interaction and – depending on the ion type - rapidly exchange with water molecules outside the shell (for alkali metal ions this exchange time – and thus the effective relaxation time of the hydration layer – is ca. $10^{-9}$ sec). (b) Simulated structure (upper figure) and density distribution of opposing hydrated lipid bilayers along the surface normal (lower figure), whereby a hydration layer (thickness of $D_w$) separates the opposing bilayers, due to the high hydration of the zwitterionic headgroups [43]. When one lipid bilayer slides pass the other, water molecules in the hydration layer lead to efficient lubrication. For panel (a), reproduced with permission from Ref. [42]; Copyright © 2012, Springer Nature.

The above-mentioned strong, short-ranged repulsion is commonly referred to as "hydration repulsion (or force)" [44], which overcomes van der Waal's attraction and so prevents direct contact of PL bilayers. Hydration repulsion is ubiquitous between hydrated species, including trapped hydrated ions, as demonstrated for hydrated sodium ions trapped between mica surfaces [45], divalent cations ($\geq 1$ M) trapped between mica surfaces [46], divalent and trivalent cations trapped between $Si_3N_4$ and sapphire surfaces [47], surfactants-coated surfaces [48–51], polymer brushes-grafted surfaces [52], as well as biological molecules, such as proteins [53].

The hydrated boundary layer also modifies the frictional force when sliding between two boundary layers takes place [54]. Raviv et. al. [45] found very low friction when compressed mica surfaces slide past each other across hydrated sodium ions trapped between them, with a frictional force comparable with the noise level even they are in close contact (inter-surface separation $D = 0.8 \pm 0.3$ nm). The extremely low frictional force arises from the fact that the bound water molecules under shear are quite fluid; indeed direct measurements indicate that the effective viscosity of the water in the primary hydration layer is only ca. 250-fold larger than that of bulk water [55]. The fluidity of confined water molecules is consistent with their rapid relaxation arising from their rapid exchange with free water molecules outside the hydration layer.

The mechanism whereby water molecules in the hydration shell bind strongly to the charge and so resist high compressive load $F_n$, while retaining their fluidity and thus experiencing low energy dissipation on shear resulting in a low friction force $F_s$ and thus a low friction coefficient, is termed hydration lubrication [42]. This applies to most hydrated species rubbing past each other, including lipid boundary layers, as confirmed by studies showing that, in addition to the exponentially decaying hydration repulsion, the friction coefficients are very low ($10^{-3} - 10^{-4}$) under pressures exceeding those in biological system [37,56]. This hydration lubrication paradigm provides insight into the origin of the very efficient lubrication of joints, and guidelines for designing more efficient boundary lubricants.

### 3. Single-component lipids

Interfacial layers of lipids have been proposed to play a key role in boundary lubrication of joints via the hydration lubrication mechanism [5]. Bilayers or vesicles of single-component PCs, which are the major components of PLs in synovial fluid, have been widely studied as a model system. Substrate-supported lipid layers are commonly prepared by Langmuir-Blodgett (L-B) methods or by direct vesicle adsorption [57]. In the L-B deposition method, bilayers are prepared by successive deposition of two leaflets on the substrate. In contrast, the morphology of layers prepared by vesicle adsorption method, which may more closely resemble the process in synovial joint [26,58], is more complex. For example, when PC vesicles are adsorbed on a mica substrate, the vesicles either remain as intact vesicles, or rupture and form bilayers, as result of the interplay between the dipole-charge interaction between PC lipids and mica and the van der Waal's interaction between hydrophobic tails [59]. Interfacial forces between PL layers strongly depend on the chemical structure of the PL, including polar headgroup and hydrophobic tails. Other factors, such as the presence of lipid molecules in the surrounding, environmental ionic strength, as well as properties of the substrate, also determines the morphology and dynamics [60] of the interfacial layer, which in turn modifies the interfacial forces.

**Alkyl chain structure and the presence of a lipid reservoir**

The alkyl chain structure of the lipid indirectly regulates the interfacial properties by determining the phase transition temperature ($T_m$) or phase state of the lipid layer via forming loosely or tightly packed hydrophobic region for fluid-state and gel-state lipid layers, respectively. Figure 3a summarizes the shear force versus normal force profiles between layers of PC lipids with different hydrophobic chain lengths. For PC in the gel-state, Goldberg et. al. [61] found that for hydrogenated soybean phosphatidylcholine (HSPC) with high phase transition temperature ($T_m = 53$ °C), gel-state small unilamellar vesicles (SUVs) adsorb on mica via dipole-charge interaction, forming a layer of closely-packed vesicles on the substrate; friction coefficients between such HSPC-SUV layers are in the range of $10^{-4} - 2 \times 10^{-5}$ up to pressures of at least 12 MPa across water. For a PC layer in the liquid-state, when 1-palmitoyl-2-oleoyl-glycero-3-phosphocholine (POPC, $T_m$ = -3 °C)-SUVs adsorbed on mica, they rupture and merge into a defect-free bilayer, exhibiting even lower friction coefficients $\mu < 10^{-4}$ under pressures up to 16 MPa across 0.3 mM POPC-SUVs, whereas the integrity of the bilayer is maintained [56]. The low friction shown by boundary PC layers is attributed to hydration lubrication at the exposed highly hydrated phosphocholine headgroups (ca. 15 water molecules per lipid, depending on how measured) [59], where the slip occurs (Figure 3b).

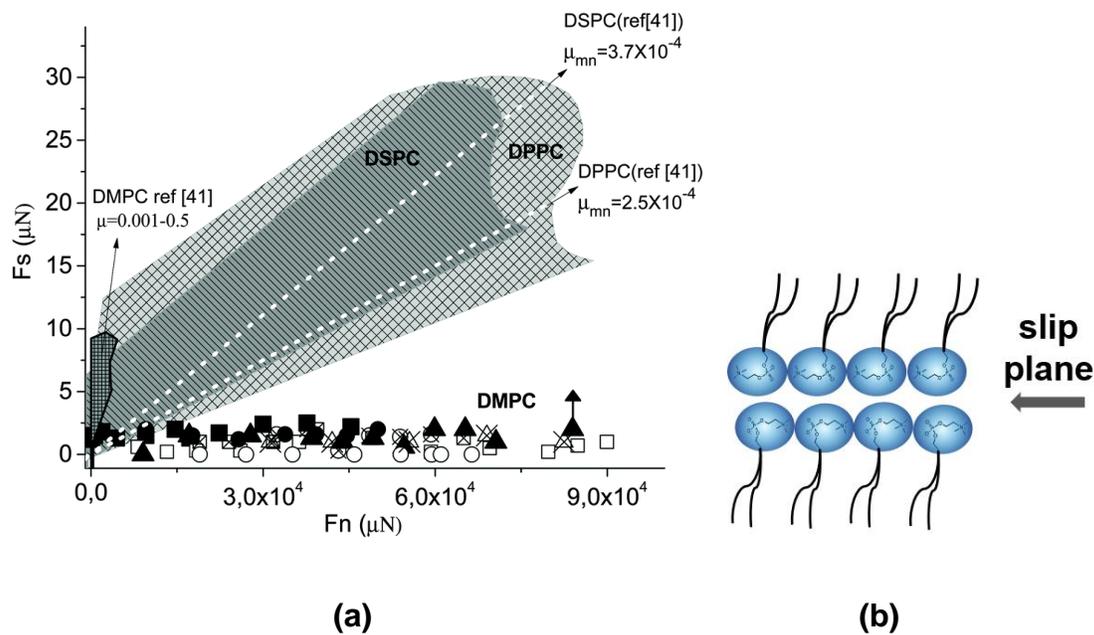

**(a)**                    **(b)**

**Figure 3.** Effects of phase state and lipid reservoir on the lubrication behavior of PC lipids. (a) Shear force versus normal force profiles for PC lipids with different alkyl chain lengths: 1,2-distearoyl-sn-glycero-3-phosphocholine (DSPC, C18:0-C18:0), 1,2-dipalmitoyl-sn-glycero-3-phosphocholine (DPPC, C16:0-C16:0), and 1,2-dimyristoyl-sn-glycero-3-phosphocholine (DMPC, C14:0-C14:0), measured across water and 0.3 mM SUV dispersion [62]. For gel-phase DSPC and DPPC interfacial layers, no significant difference in the friction coefficients is observed whether under water (dashed lines) or across SUV dispersions (shaded areas). However, for liquid-state DMPC layer, friction coefficients measured across water (left darker shaded area) is much higher compared with those in the presence of reservoir (solid and empty symbols). (b) Slip-plane for the opposing PC layers is at the highly hydrated headgroup-headgroup interface [63]. Panel (a) is reproduced with permission from Ref. [64]; Copyright © 2013, Royal Society of Chemistry. Panel (b) is reproduced with permission from Ref. [63]; Copyright © 2015, American Chemical Society.

Figure 3a also shows different lubrication behavior for PC liquids under conditions with and without reservoir. Gel-state PC lipids, HSPC and 1,2-dipalmitoyl-sn-glycero-3-phosphocholine (DPPC), are more closely packed in the bilayer and robust enough to sustain higher compression meanwhile maintaining the integrity of the interfacial layer even without reservoir. As a result, no significant difference is observed. However, when measurements are performed for the 1,2-dimyristoyl-sn-glycero-3-phosphocholine (DMPC) system across water, friction force increase abruptly at higher loads, probably because of degradation and rupture of the liquid-phase DMPC bilayer structure under load and shear [59,65]. In contrast, in the presence of a DMPC lipid reservoir, as seen for measurements in the DMPC-SUV dispersion (Figure 3a), the friction remains very low even up to high contact pressures [62], probably due to the ease of defect healing of this liquid-phase lipid in the presence of free lipids. Such defect healing is further confirmed by higher friction coefficients accompanied by introducing defects to the bilayer through forcing the liquid-state interfacial layer pass air-water interfaces [59], infinite dilution of the bulk liquid-state SUV solution [63], or depositing PL layer at a higher temperature (bilayer in liquid-state) followed by cooling down [66].

**Effect of different PL headgroups**

In addition to PC lipids, zwitterionic SMs and PEs, as well as negatively charged PLs, PA and PG, have also been identified from synovial joints [30,67]. PL headgroups directly determine the properties of the interfacial layer, including surface charge, dipole potential, hydration level, as well as interaction with the matrix [68]. Unlike PCs, where there is no direct hydrogen-bonding between the phosphocholine headgroups, intramolecular hydrogen-bonds form between the phosphate and hydroxyl groups of SM molecules [69], and for PE lipids, inter-/intra-molecular hydrogen-bonds form between the amine and phosphate/carbonyl group of their headgroups [70]. As a result, phase transition temperature for fully hydrated PLs which have the same alkyl chain structure (C16-C16) but different headgroups follows the order of 1,2-dipalmitoyl-sn-glycero-3-phosphoethanolamine (DPPE, 63.0 ºC) > N-palmitoyl-D-erythro-sphingosylphosphorylcholine (PSM, 40.7 ºC) ≈ DPPC (40.7 ºC) ≈ 1,2-dipalmitoyl-sn-glycero-3-phospho-(1'-rac-glycerol) (DPPG, 40.6 ºC) [63,71,72]. This is relevant for lubricating boundary layers as the more robust bilayers (those with higher $T_m$) can in principle withstand higher compressive and shear stresses.

The friction coefficients between these lipid layers are in the order of DPPC ≤ egg SM < DPPE < DPPG-$Ca^{2+}$, ranging from $\mu \approx 10^{-4}$ for DPPC and egg SM under pressures of more than 10 MPa to $10^{-2}$ for DPPG-$Ca^{2+}$ under ca. 1 MPa (unpublished data). All these major components of lipids in synovial joints by themselves are thus capable of efficient reduction of sliding friction, attributed to the hydration lubrication mechanism.

**Effect of electrolytes**

Compared with friction coefficients measured between PC layers across water, those between PC layers immersed in physiologically high salt concentrations (150 mM $NaNO_3$) are relatively higher (Table 1). It clearly indicates that, the friction coefficients obtained in physiologically high salt concentration are consistently higher than those obtained across water. A possible explanation is that high concentration monovalent ions compete with lipid molecules for hydration water, resulting in less hydrated PL molecules and higher friction coefficients. This is also in line with early studies on friction between phosphocholinated brushes [52]. Nonetheless, boundary layers of HSPC [61], egg SM [73], or DPPC [19] are still capable of affording a level of lubrication comparable to or better than that of articular cartilage under physiologically high salt concentrations and pressures.

**Table 1**. Comparison of friction coefficients of boundary PL layer measured across water and across physiologically-high concentration salt.

| System | Across water | Across physiologically-high concentration salt | Ref. |
|---|---|---|---|
| HSPC-SUVs on mica | $\mu \approx 10^{-4} - 2\times 10^{-5}$ <br> $P \approx 12$ MPa | $\mu \approx 0.008 - 0.0006$ <br> $P \approx 6$ MPa | [37,61] |
| egg SM-SUVs on mica | $\mu \approx (5.3 \pm 0.8) \times 10^{-4}$ <br> $P \approx 9.8 \pm 0.2$ MPa | $\mu \approx (2.0 \pm 0.8) \times 10^{-3}$ <br> $P \approx 7.6 \pm 0.8$ MPa | [73] |
| DPPC-SUVs on HA-coated substrate | $\mu \approx (1.5 \pm 1) \times 10^{-3}$ <br> $P \approx (12.2 - 22.0)$ MPa | $\mu \approx 7 \times 10^{-3}$ <br> $P \approx 16$ MPa | [19] |

A further effect of high salt concentration is that under physiologically high salt conditions (150 mM NaNO$_3$), the Debye length for electrostatic interactions is ca. 0.7 nm; this screens electrostatic repulsions, as between like charges, and facilitates the adsorption of weakly charged lipid vesicles on the substrate of the same charge [73,74]. Moreover, in addition to monovalent cations, synovial fluid also contains lower concentrations (ca. 2 mM) of divalent cations, including calcium, magnesium, and zinc [75]. Such cations also act as bridges binding negatively charged lipid layers to a negatively-charged substrate (we recall that articular cartilage is negatively charged [76]). The interaction between divalent cations and the lipid bilayers themselves is more complex, and strongly depends on the nature of PL. Briefly, calcium ions interact more strongly with negatively charged lipids than with neutral ones, and bind more strongly to gel-state bilayers such as DPPC than to fluid-state ones such as POPC, resulting in more rigid lipid bilayers and dehydrated lipids [77].

**Effect of substrate**

Boundary layers on the cartilage surface, which have been proposed to consist of a synergistic complex of HA, lubricin and PLs, are responsible for the very efficient boundary lubrication of articular joints. Here HA is considered to be complexed with interfacial PL layers [5]. Therefore, HA-coated surfaces, either by chemical interactions or physical adsorption, have been widely utilized to model the negative-charge and configuration of biomacromolecules at the cartilage surface. Seror and Zhu et. al. [20,21] found that the friction coefficients between a surface-coated HA mica substrate complexed with HSPC- or DPPC-SUVs are as low as $\mu \approx 0.001$ at the highest applied contact pressures up to 10 MPa, far lower than for surface-attached HA alone, for which $\mu \approx 0.3$. Meanwhile, cryo scanning electron microscopy (cryo-SEM) images clearly show "hollowed-out" HSPC-SUVs or DPPC-SUVs forming a thread-like structure on the HA polymers extending from the surface, quite different to the morphology of closely-packed vesicles on bare mica substrate (Figure 4).

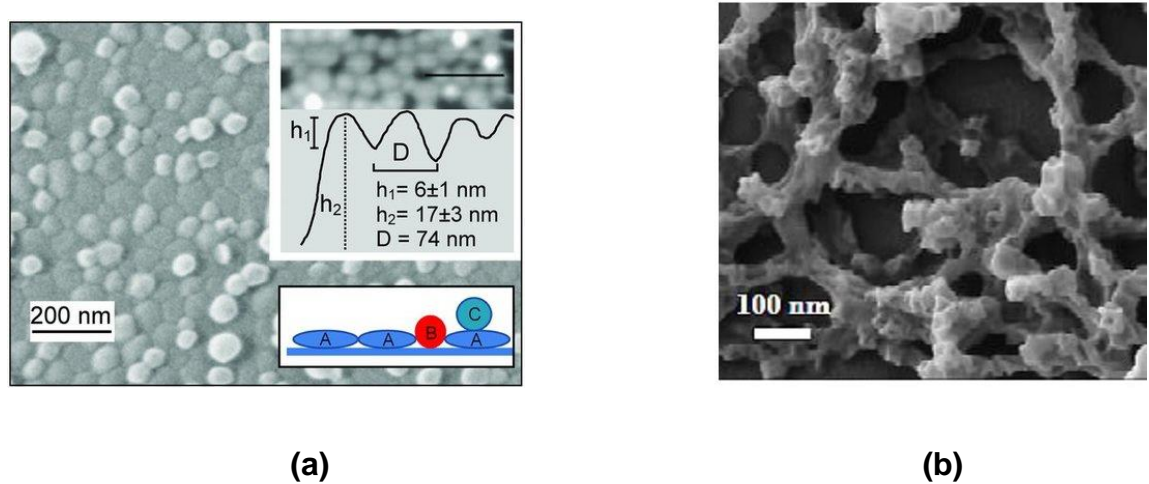

**(a)**                                                    **(b)**

**Figure 4**. Morphologies of HSPC-SUVs adsorbed on (a) mica [37]; and (b) avidin-biotinylated HA-coated mica under water [21]. A closely-packed layer of HSPC-SUVs was observed on mica substrate. On the HA-modified substrate, the HSPC-SUVs (which are "hollowed-out" by the cryo-SEM protocol) attach to the HA polymers, forming a thread-like structure. Panel (a) is reproduced from Ref. [37]; Copyright © 2011 WILEY-VCH Verlag GmbH & Co. KGaA, Weinheim. Panel (b) is reproduced from Ref. [21]; Copyright © 2017 Acta Materialia Inc. Published by Elsevier Ltd.

Consistent with this is the reduced friction achieved by HA-HSPC boundary layers on different substrates, including intrasynovial tendon surfaces, cartilage-mica or cartilage-glass interfaces, and between silica substrates [10,78,79]. However, when DMPC- or POPC-SUVs are complexed with HA in lipid-free aqueous medium, such layers readily rupture, likely due to the liquid-state of the lipids, and provide much poorer lubrication [80]. Similarly, and for likely the same reason, interfacial gel-state PC (HSPC, DPPC)-SUV layers, when adsorbed onto (chitosan-alginate)- or onto poly(ethylene oxide)-coated substrates also exhibited friction coefficients at least one order of magnitude lower than those for fluid-state DMPC bilayers formed by rupture of SUVs [81,82].

In addition to the nature of PCs, the molecular weight of underlying HA molecules also determines the morphology of HA-PC complex and further regulate the lubrication efficiency [83]. Zhang et. al. [17] found that median- and low-molecular weight HA on a gelatin substrate complexed with HSPC-SUVs cannot provide as good lubrication as high-molecular weight HA. This was attributed to the fact that the weaker overall adhesion between lower-molecular weight HA and the underlying gelatin layer causes the HA-PC complex to be removed from substrate during shear. These results may be related to the fact that osteoarthritic joints, where lubrication is taken to be less effective than in healthy joints, are associated with decreased molecular weight of HA relative to healthy joints [84]; they may indicate the benefit of injecting high molecular weight HA for OA treatment [85].

More recently, Lin et. al. [86], inspired by the proposed lipid-based boundary lubrication mechanism of articular cartilage, which may be viewed as a complex biological hydrogel, designed PC-lipid-incorporating hydrogels (of poly(hydroxyethylmethacrylate), pHEMA, and other hydrogels. They discovered that the friction coefficient was reduced by up to 2 orders of magnitude (in the case of pHEMA, to values $\mu \approx 0.02$ to 0.005) relative to the lipid-free hydrogels. This was shown to arise from lipid-based boundary layers at the gel surface which as they wore away were continuously replenished by lipid-microreservoirs within the bulk of the gel. Both friction and wear were thus massively reduced as compared to the lipid-free hydrogels.

Overall, these studies, whether using nanotribometry such as with surface force balances, or at macroscopic level with a standard tribometer clearly show that the PC-lipid-based boundary layers can provide extreme reduction of sliding friction. This occurs through the hydration lubrication mechanism acting at the highly-hydrated phosphocholine headgroups exposed by the lipid assemblies. For single-component lipids, the efficiency of friction-reduction is related to the phase state: boundary layers consisting of gel-state PC-SUVs are more robust, and thus better lubricants at high pressure, than those of fluid-state PC bilayers.

## 4. Lipid mixtures

In synovial joints, both in the synovial fluid and within the cartilage, complex mixtures of lipids with different headgroups and alkyl chain structures have been identified [27,28]. The mixture is composed of both neutral and negatively-charged lipids, and most of its PL components have unsaturated alkyl tails, and are thus expected to be more fluid-like than gel-like at physiological temperatures. Such lipids may assemble on the cartilage surface in forms of multilamellar vesicles or lipid layers, as indicated by electron microscopy [87,88], or complexed with other macromolecules, and their proposed role in the cartilage boundary lubrication has been noted above.

Compared with single-component lipids, whose assembly in boundary layers is in either the gel- or the liquid-phase and is rarely in the so-called ripple-phase (where gel- and liquid-phases coexist [89]), lipid mixtures exhibit more complicated phase behavior. These may manifest as lateral phase separation caused by the immiscibility between different PL molecules [90], dynamic and heterogeneous distribution of PL in the membrane [91], and asymmetric distribution of PLs across the membrane (i.e. in its two leaflets) due to the configurational characteristics of the lipids [92,93]. It may also be associated with a more sensitive response to changes in environmental conditions [94]. Therefore, the morphology of lipid boundary layers and the interactions between opposing layers may be considerably more complicated for mixed lipids than for single-component ones.

**Hydration lubrication, hemifusion, and self-healing**

Normal force profiles between boundary layers on mica, composed of lipid mixtures, such as PC -cholesterol and binary saturated-unsaturated PC lipids (DPPC-POPC), show short-ranged hydration repulsion similarly to single-component layers [62], as shown in Figure 5. However, above a certain pressure, the separation between the two mica substrates dropped from ca. 10 nm – corresponding to two bilayers – to 5 nm, corresponding to a single bilayer, indicating hemifusion of the trapped bilayers [63,95]. Hemifusion was observed for gel-state PCs, DSPC and DPPC, incorporating with 40% cholesterol, which modulates the fluidity of the bilayers, and also for DPPC-POPC mixtures (molar ratios 8:2, 5:5, and 2:8). This is attributed to the height mismatch, or defects in the interfacial layer, arising from the different lipid sizes, which lowers the energy for hemifusion by exposing hydrophobic moieties at the interface at high compression, thus triggering the process via hydrophobic attraction between the opposing exposed tails [96].

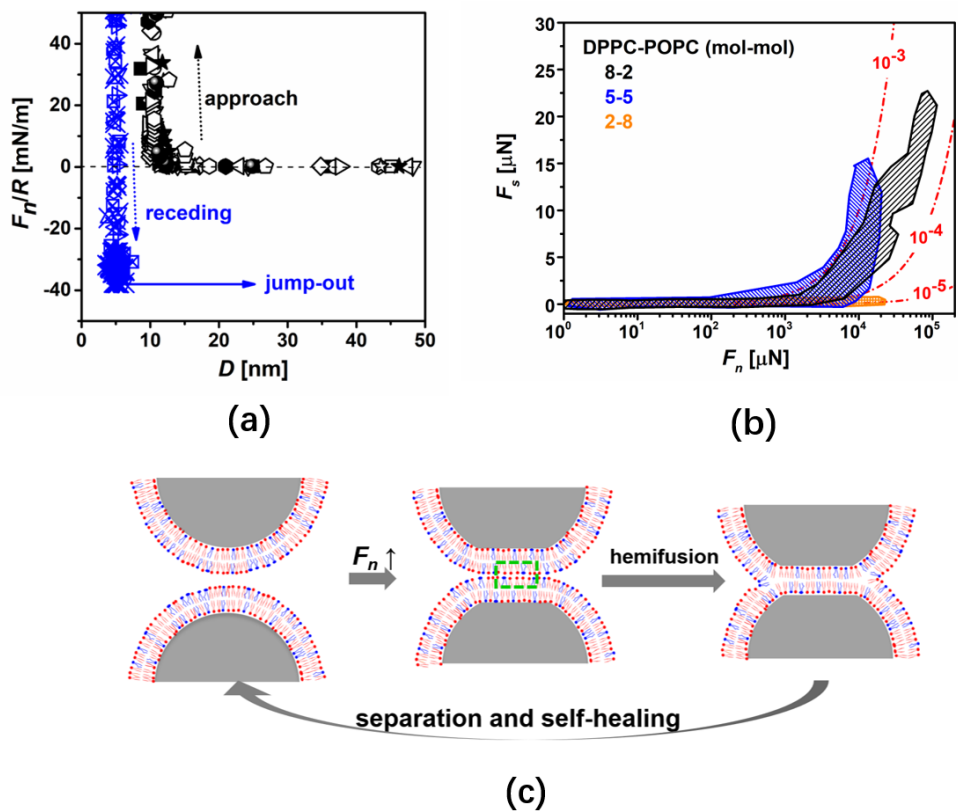

**Figure 5**. Boundary lubrication, hemifusion, and self-healing of DPPC-POPC mixtures. (a) Representative normal force profiles for the DPPC-POPC (2-8, molar ratio) mixture deposited on mica. Hemifusion is indicated as a shift in the separation distance from ca. 10 nm at approach to 5 nm upon receding (where $D = 0$ corresponds to mica-mica contact), and a jump-out when separating two surfaces. (b) Shear force versus normal force profiles show that all the three mixtures exhibit good boundary lubrication, coefficients varying between $10^{-3} - 10^{-5}$ before hemifusion. (c) a cartoon illustrating the lubrication, hemifusion, and recovery (self-healing) of bilayers upon separation. Adapted from Ref. [63]. Copyright 2015, American Chemical Society.

Prior to hemifusion, friction coefficients between the interfacial PC mixtures are in the range $10^{-2}$ to $10^{-4}$ for PC-40% cholesterol and are $< 10^{-3}$ for DPPC-POPC mixtures [63,95]. Such low friction coefficients are also attributed to hydration lubrication at the slip plane between the opposing exposed, hydrated zwitterionic headgroup layers. However, following hemifusion, the lowest-friction interface disappears, resulting in an abrupt increase in friction force for the studied system, with a friction coefficient $\mu >$ ca. 0.01. The slip-plane shifts to headgroup-mica or tail-tail interface at hemifusion [74], both of which involve much higher energy dissipation on sliding, due to breaking and reforming of van der Waals bonds (in the case of slip at the tail-tail interface) or charge-dipole bonds (in the case of slip at the head-group-mica interface).

Hemifusion of a compressed lipid bilayer has also been detected for a phase-separated DOPC/brain-SM/cholesterol mixture [97], biomimicking PC-PC-SM- phosphatidylserine (PS)-cholesterol mixtures [98], as well as lipid mixtures extracted from healthy or osteoarthritic human synovial fluid [74] and lipid mixtures mimicking those in synovial joints [99]. As hemifusion is accompanied by increased shear force and damaged interfacial layers upon sliding, prevention of hemifusion or suppressing it to higher normal loads/pressures could be beneficial to the lubrication facilitated by interfacial layers of PL mixture.

**Synergistic effects**

The robust gel-phase and fluid liquid-phase lipids plays different roles in the lubrication. Previous results suggested that, when phase-separated gel- and liquid-state patches coexist on the boundary layer, such as for the case of DPPC-POPC in a molar ratio of 8:2, the opposing gel-state patches are capable of bearing higher normal load and resist hemifusion, leading to partial hemifusion between liquid-state patches[63]. However, when the hemifused interfacial layers self-heal into two bilayers following their separation, aided by lipids in the reservoir, the process is quite rapid for fluid-state lipid molecules, as the bilayer is more fluid and can accommodate new lipids, but can take up to days when gel-state lipids are involved due to the rigidity of the bilayer [56][100]. Therefore, we propose that gel- and fluid-state lipids acting together may contribute to efficient friction-reduction boundary layers in different ways: The gel-state lipids are more robust and thus responsible for low friction up to high pressures, while the fluid-state lipids present in the surrounding dispersion can more easily healing the boundary layer following any damage due to hemifusion.

Finally, we constructed PC-SM-PE-PA mixtures whose composition emulates that of the unsaturation and charge characteristics of lipids found in synovial joints, and examined their lubrication properties as supported boundary layers. Our results [99] show that in presence of physiological concentrations of monovalent and divalent salts, such mixtures have a clear synergistic effect on the stability and lubricating ability of the lipid-mixture layer, highlighting the synergistic effect that lipid mixtures may have for biolubrication when coating articular cartilage in joints.

## 5. Conclusion and perspectives

Boundary layers comprising single-component or mixed lipids identified in synovial joint are capable of providing efficient lubrication and pressure-bearing through the hydration lubrication mechanism at the slip-plane between the hydrated headgroups exposed at their outer surface. In this review, we summarize factors influencing the boundary lubrication properties of single- or multi-component lipids, including the chemical structure of lipid(s), the presence of lipid reservoir, environmental electrolytes, molecules on the substrate, as well as the composition and ratio of lipids in a mixture. Improved boundary lubrication ability may be attained by fine-tuning these factors, with possible implications for lubrication of biomedical devices (e.g., hydrogel-based devices such as contact lenses or catheter coatings) or for the treatment of early-state OA via intra-articular injection of suitable lipid-based lubricants.

Progress in understanding the lubrication ability of lipid-based boundary layers points to some outstanding questions. These include: (1) Is there a correlation between enhanced lipids concentration and composition that have been observed in the synovial fluid of OA patients, and the lubrication ability of the articular cartilage? (2) Is there a dynamic exchange between lipids in the synovial fluid and those on the cartilage surface? If so, how might this affect the boundary lubrication of cartilage? Answering these, which will require both lipidomic analysis of joint lipids and its correlation with cartilage lubrication, will provide deeper insight into the relation between lipids, cartilage lubrication, and joint homeostasis.

**Conflict of interest statement**

The Weizmann Institute has patents and pending patent applications on the use of liposomes as lubricants.

**Acknowledgements**

The authors thank the European Research Council (CartiLube, Grant No. 743016), the Israel Ministry of Science and Technology (Grant NO. 713272), the Israel Science

Foundation-National Natural Science Foundation of China (ISF-NSFC) joint program (Grant No. 2577/17), the McCutchen Foundation, and the Israel Science Foundation (Grant No. 1229/20) for their support of some of the work described in this review. This work was made possible in part by the historic generosity of the Harold Perlman family.

PLs are assymetrically and dynamically distributed across the Gram-negative bacteria membrane.